\documentclass[
reprint,
 amsmath,amssymb,
 aps,
prfluids,
]{revtex4-2}
\usepackage{xcolor}
\usepackage{comment}
\usepackage{booktabs}
\usepackage{multirow}
\usepackage{graphicx}
\usepackage{appendix}
\usepackage{dcolumn}
\usepackage{bm}
\usepackage{xcolor}

\def\tb{\textcolor{blue}}

\usepackage{ulem}
\usepackage{multirow}

\begin{document}


\preprint{APS/123-QED}

\title{Static and Dynamic Estimation of Flexural Rigidity of Soybean}

\author{Akihito Kiyama}
\affiliation{Department of Biological and Environmental Engineering, CALS, Cornell University, USA}
\affiliation{Graduate School of Science and Engineering, Saitama University, Japan}

\author{Haruka Tomobe}
\affiliation{Department of Civil and Environmental Engineering, Tokyo Institute of Technology, Japan}

\author{Jisoo Yuk}
\affiliation{Department of Biological and Environmental Engineering, CALS, Cornell University, USA}

\author{Sharon Wang}
\affiliation{Department of Biological and Environmental Engineering, CALS, Cornell University, USA}

\author{Sunghwan Jung}
\affiliation{Department of Biological and Environmental Engineering, CALS, Cornell University, USA}
\email{sj737@cornell.edu}

\date{\today}

\begin{abstract}
This paper evaluates a cantilever system as a simple method for measuring the mechanical rigidity (bending rigidity $EI$ or Young’s modulus $E$) of plants. Using soybeans as a test sample—whose $E$ values have rarely been reported—we conducted static and dynamic experimental measurements alongside numerical modal analysis. Results showed good agreement in $EI$ values between the static and dynamic tests when branches and leaves were removed, provided the stem responded uniformly. However, the added mass of attached foliage causes complex dynamic interactions, which we analyze through both experimental and numerical approaches. Ultimately, our findings suggest that cantilever force-deflection measurements provide a practical, on-site approach, contributing to the development of an affordable and reliable standard for estimating plant mechanical properties.   
\end{abstract}

\keywords{Plant rigidity, Vibration, Soybean}
\maketitle

\section{Introduction}

Lodging caused by stem bending (e.g., wind-induced lodging) is one of the major problems in agriculture~\cite{Berry2004,Pinera-Chavez2020}. There are two primary types of lodging (stem lodging~\cite{Baker1998,Ookawa2014} and root lodging~\cite{Brune2018,Tomobe2019}), both of which initiate with the temporary or permanent bending of aboveground structures. To understand how plants respond to these bending forces, it is crucial to characterize their mechanical properties~\cite{gart2015droplet, bhosale2020bending, yuk2022visual, jung2021measuring, fuchs2021fluttering, yuk2026leaf}. To predict stem bending, researchers frequently estimate Young's modulus, $E$~\cite{Berry2003, AlZube2018, Nakashima2023}, which provides a fundamental material property representing the intrinsic rigidity of the tissue~\cite{Baker1995}. Within the limit of elastic deformation, Young's modulus is a useful metric for quantifying the rigidity of plant materials. The Young's modulus is directly determined with the tissue-scale structure  and lignin concentration~\cite{Ookawa2014}. Therefore, in plant stems, a high Young's modulus indicates that the stem's microstructure and chemical composition are mechanically resilient. A low Young's modulus in the aboveground structure can also cause root lodging. In structural engineering, it has been known for decades that lateral deformation of above-ground structures can generate large moments in the entire structure, including its underground components; this phenomenon is known as the P - $\Delta$ effect~\cite{Gaiotti1989,Gupta2000,Bird2023}. In plants as well, it can be inferred that a low Young's modulus in the aboveground parts results in an excessive moment in the underground parts due to the P - $\Delta$ effect. Therefore, the Young's modulus of the aboveground parts of the plant, particularly the stem as its main structural component, is thought to directly influence the risk of both stem lodging and root lodging.

Several studies have attempted to estimate crop rigidity, with maize being one of the most extensively studied~\cite{VonForell2015, Huang2016, AlZube2018, Shu2020, Berry2021, Ottesen2023, Oduntan2024}. Methods range from finite element analysis relating tissue-scale Young's modulus to macro-scale bending strength~\cite{VonForell2015}, to conventional 3-point bending tests yielding comparable values ($E\sim \mathcal{O}(10)$~GPa)~\cite{AlZube2018}. In contrast to conventional destructive testing, non-destructive and invasive alternatives have also been developed. For example, Niklas and Moon~\cite{Niklas1988} utilized a multiple resonance frequency spectra method for {\it Allium sativum L} (i.e., garlic), while Nakashima et al.~\cite{Nakashima2023} employed stem shear-wave detection. Furthermore, a recent vibration-based approach by Ogilvie and Cook~\cite{Ogilvie2025} demonstrated that the presence of leaves directly affects the natural frequencies of maize plants.

Despite these advancements, significant research gaps remain. First, as highlighted by Shah et al.~\cite{Shah2016}, there is still no standard, widely accepted method for measuring the mechanical properties of plants at the macroscopic scale, particularly for on-site applications. Second, while prior research has successfully characterized simple, single-stalk crops, the dynamic deformation of crops with complex, branched structures such as soybeans ({\it Glycine max} (L.) {\it Merr.}) and buckwheat ({\it Fagopyrum esculentum}) remains poorly understood.

To address these gaps, this paper tests a simple, affordable cantilever beam system to estimate the rigidity of branched plants (i.e., soybean). Based on classic beam theory and commonly used as the standard test for wood poles, this system is highly practical for non-destructive, in-pot measurements~\cite{Caliaro2013}. We focus on the soybean due to its global agricultural importance for fuel, feed, and plant-based protein, as well as its high susceptibility to lodging~\cite{Piper1867, Umburanas2022, Yamaguchi2014, Li2024}. It is known that, even when the yield per unit area of soybeans changes, the harvest index which is the proportion of edible parts relative to total biomass remains relatively constant~\cite{Kawasaki2016}. Therefore, securing a larger above-ground biomass is necessary to increase soybean yield per unit area, and historically, preventing lodging has been crucial to achieving this~\cite{Piper1867}.

After validating our methodology with a homogeneous polycarbonate beam, we employ both force-deflection (static) and free vibration (dynamic) tests on actual soybean samples, systematically varying the location and direction of the applied forces. 
We estimated the flexural rigidity, $EI$, for branched soybean plant. This $EI$ value is a product of the Young's modulus and the second moment of area, which represents the structural rigidity to the bending force. 
We also discussed the influence of the branched structure, especially to the accuracy of the dynamics test, with a help of numerical simulation. It poses a limitation of the dynamic vibration test approach, where the details of the plant structure is expected to be not available. Further, we discussed conversion from estimated $EI$ to $E$ values, which shows directions for future studies. 

\section{Materials and Methods}
\subsection{Materials}
We used two different material types, as summarized in Table \ref{tab:material}. First, a poly-carbonate beam was used for validation purposes. Its Young's modulus is generally known to be approximately $E\sim2-3\times10^9$~Pa, which falls within a similar range to that of plants~\cite{Oduntan2024}. We varied the experimental conditions for both the static and dynamic tests (see the Methods section below for details). Next, we tested soybean plants grown in the laboratory for approximately one month. 

Representative images of the soybean plants are shown in figure~\ref{fig:setup}(a). Note that the plant in figure~\ref{fig:setup}(a) was grown from the same seed batch and in a similar age and morphology with other plants, while this particular sample was not subjected to tests. The typical height of the soybean plants were approximately 20 to 30~cm, with a few flexible branches and leaves. We note that the leaves and branches were heavier than the main stem (see table~\ref{tab:material}, specifically the values in parentheses). 
We first tested both static and dynamic characteristics of the soybean plant immediately after cut from the pot, while keeping the branches and leaves remain connected. We then removed the branches and leaves from the stem, and performed the same measurements to isolate the mechanical influence of the foliage. Detailed structural dimensions and masses for all samples are listed in Table \ref{tab:material}. We note that a soybean plant (marked by sample \# 2) was subjected to another test, where the stem was rotated approximately 90$^\circ$ to test its influence on the macroscopic behavior.


\begin{table}
    \centering
    \begin{tabular}{cllll}
        \toprule
        Group & sample \# & $L$~(cm) &$L_\mathrm{net.}$~(cm) & $M$~(g) \\
        \midrule
        Polycarbonate$^{\ast1}$            & 1 & 25 & 12.5 - 25 & 7.36 \\
        \midrule
        \multirow{3}{*}{Soybean} & 1 & {21} & {5 - 21} & {1.29 (4.33)}$^{\ast2}$ \\
                                 & 2 & {23} & {10 - 23} & {1.77 (7.83)}$^{\ast2}$ \\
                                 & 3 & {29} & {4 - 29} & {1.59 (6.02)}$^{\ast2}$ \\
                                 & {4} & {{22}} & {5 - 22} & {1.14 (3.41)}$^{\ast2}$ \\
        \midrule
        \bottomrule
    \end{tabular}
    \caption{Materials used in the experiments. $L_\mathrm{net}$ represents the measurement condition for the static (force-deflection) test (see the Methods section). We repeated the measurement at least two times for each condition. $^{\ast1}$ The poly-carbonate beam has a rectangular cross-section, where the width $b$ and thickness $h$ of the cross section were respectively 20~mm and 9~mm$^{\ast2}$. The numbers in parentheses indicate the mass of the whole plant with branches and leaves, whereas $M$ represents the mass of the stem. } 
    \label{tab:material}
\end{table}

\subsection{Static and Dynamic Estimation of Flexural Rigidity of Soybean}
As mentioned, we employ two complementary experimental methods to estimate the rigidity of the soybean stem, detailed below.

\subsubsection{Static test}
\begin{figure}
    \centering
\includegraphics[width=0.95\columnwidth]{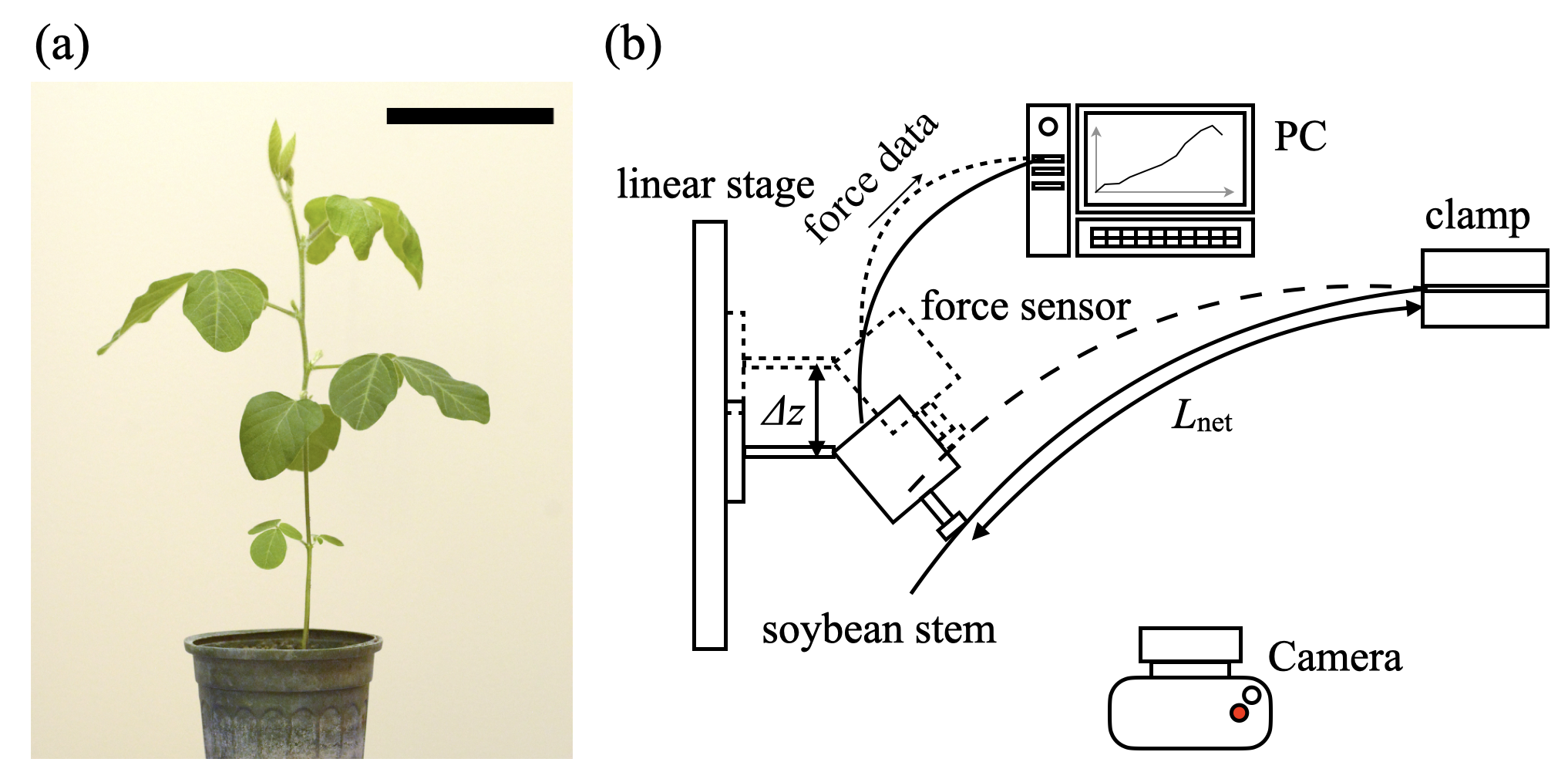}
    \caption{(a) A representative soybean plant showing the typical characteristics of the specimens tested. (b) An illustration of the experimental setup. A clamp fixes one end of the stem, while an indenter mounted on a force sensor pushes the other end at $L_\mathrm{net}$. Force data and camera images are acquired simultaneously. In the dynamic (vibration) test, free vibration is induced by flicking the stem with a finger.}
    \label{fig:setup}
\end{figure}

The static test was conducted based on a standard cantilever bending configuration~\cite{Tomobe2019}. The one side of the sample (i.e., the root-side stem) is clamped horizontally to fix the overall structure. 
During the static test, we applied the force $\Delta F$ to the clamped stem using a small indenter mounted on a force sensor, while monitoring the deflection $\Delta z$. Treating the cantilever stem under classical beam theory allows us to estimate the relationship between the measured force $\Delta F$ and the deflection $\Delta z$ as
\begin{equation} \label{eq:force-deflection}
    \Delta z=\frac{L_\mathrm{net}^3}{3EI}\Delta F, 
\end{equation}
where $E, I$ and $L_\mathrm{net}$ are the Young's modulus, the second moment of area, and the net length of the stem, respectively. The net length $L_\mathrm{net}$ represents the distance from the clamp to the loading point (see Figure \ref{fig:setup}(b)). 

This equation can be reformulated for the static bending stiffness as
\begin{equation} \label{static}
    (EI)^\mathrm{static}=\frac{\Delta FL_\mathrm{net}^3}{3\Delta z},
\end{equation}
which allows us to determine the flexural rigidity obtained by the static test, $(EI)^\mathrm{static}$, from a relation between the measured force $\Delta F$ and displacement $\Delta z$.

In the experiment, we employed a calibrated force sensor (FUTEK, USB 220) mounted on a motor-controlled linear stage (VELMEX BiSlide). The displacement of the sensor, the temporal location of the stem (i.e., the deflection $\Delta z$ of the forced point at $L_\mathrm{net}$) was measured through images taken by a digital single-lens reflex (DSLR) camera. We used ImageJ software (National Institutes of Health) to measure $\Delta z$ from the images, where a typical spatial resolution was about 20~pixels/mm. The linear correlation between $\Delta F$ and $\Delta z$ was calculated via least-squares linear fitting. Note that this method estimates the ``macroscopic" bending behavior of the stem under the assumption that the structural properties of the stem (specifically its density and rigidity) are uniformly distributed along its body.

\begin{figure}
    \centering
\includegraphics[width=0.95\columnwidth]{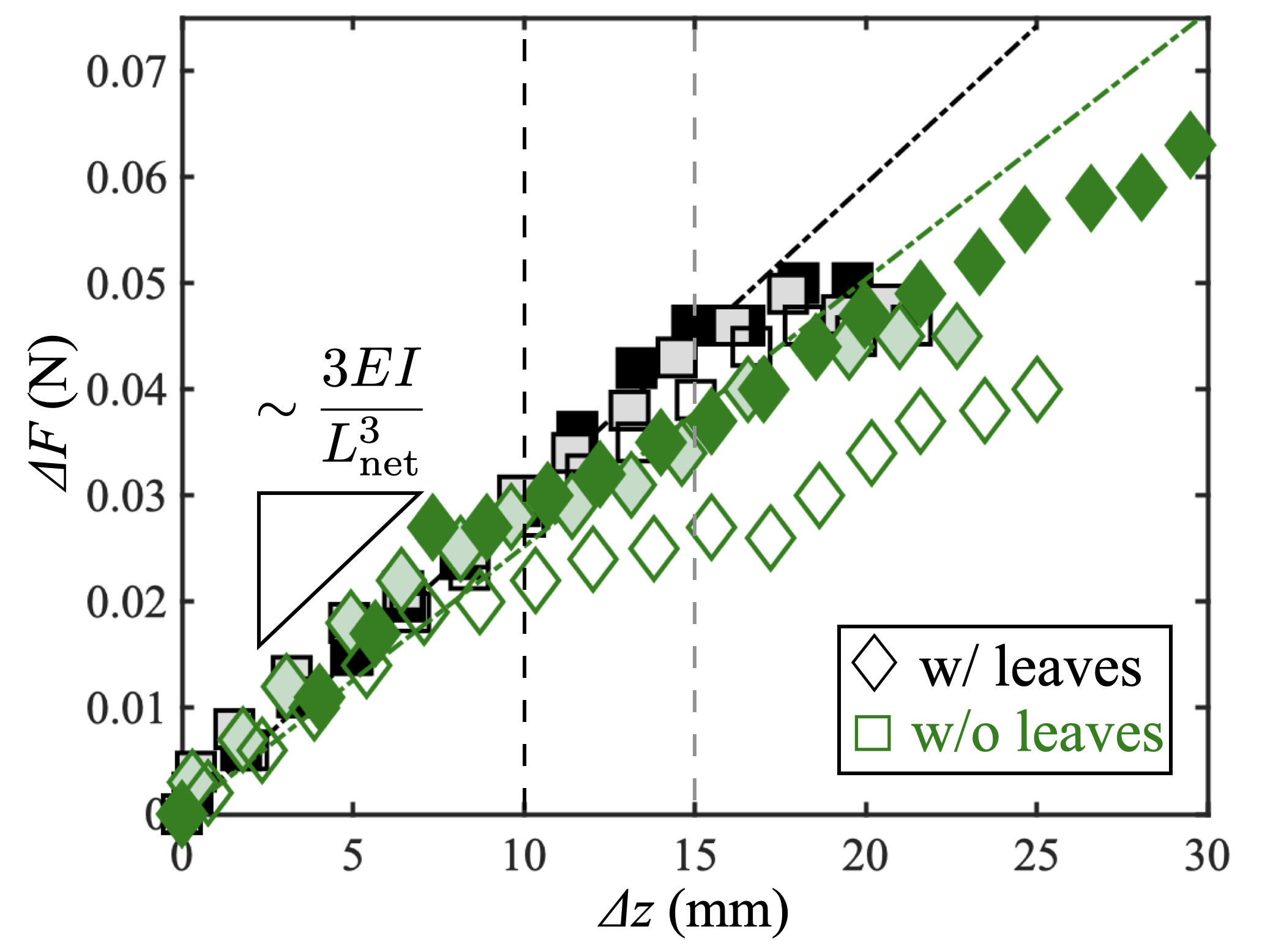}
    \caption{Force-deflection ($\Delta F$ vs. $\Delta z$) relationship for a soybean stem loaded at $L_\mathrm{net}=17$ cm. Green and black markers represent data for the stem with and without branches and leaves, respectively, with marker colors distinguishing individual trials. Dashed lines indicate linear fits applied to data up to $\Delta z < 15$ mm to illustrate general trends. While the linear relationship deviates at larger deflections, data within $\Delta z < 10$ mm demonstrate a robust linear response and were used to calculate the static bending stiffness (slope = $3EI/L^3_\mathrm{net}$).}
    \label{fig:forcemeasurement}
\end{figure}

To describe our data processing routine, Figure~\ref{fig:forcemeasurement} shows a typical $\Delta F - \Delta z$ plot, where green and black markers show the data for the stem with and without branches and leaves, respectively. The face color of the markers distinguishes the individual trials (three trials for each condition). As discussed earlier (see Equation \ref{eq:force-deflection}), the slope indicates the $EI$ value as long as the infinitesimal deformation assumption is valid. The black and green dashed lines show linear fits for each data series, in which the fits were applied for data up to $\Delta z<15$~mm to show general trends. It is obvious that the green data exhibit a smaller slope than the black data. It may be attributed to not only unstable (slippery) contact between the sensor and the stem but also large deflections. During the analysis, we monitored the camera images to ensure that the sensor remained in contact with the stem. 
In general, the deviations become more pronounced as $\Delta z$ increases.

In Figure~\ref{fig:forcemeasurement}, the data seem to less deviate around $\Delta z<10$~mm, which is approximately 6\% of the net length $L_\mathrm{net}$ for this case. Based on these preliminary considerations, we restricted our analysis to data satisfying $\Delta z/L_\mathrm{net}\leq5\%$ in the later sections. We consistently maintained this routine when estimating $(EI)^\mathrm{static}$ from the static measurements.

\subsubsection{Dynamic test}
After the static test, we then gently sway the stem with a finger to initiate free vibration. In this test, we set the point of the finger excitation at the same location as $L_\mathrm{net}$ in the static test for consistency. For the uniform (homogeneous) cantilever beam, Euler-Bernoulli beam theory predicts that the first mode of the fundamental frequencies is given by
\begin{equation} \label{eq:omega}
    \omega_0=2\pi f_0=3.52\sqrt{\frac{EI}{ML^3}},
\end{equation}
where $M$ is the mass of the clamped stem excluding a portion that was sandwiched by clamp, which was measured separately using a weight balance. Once we obtain the frequency of the freely vibrating stem, we can estimate the rigidity as
\begin{equation} \label{dynamic}
    (EI)^\mathrm{dynamic}=\bigg(\frac{\omega_0}{3.52}\bigg)^2~ML^3.
\end{equation}
We note that the characteristic length of the stem \tb{$L$} is taken as the distance from the stem tip to the clamp, regardless of the location where the perturbation is applied. 

\begin{figure}
    \centering
\includegraphics[width=0.95\columnwidth]{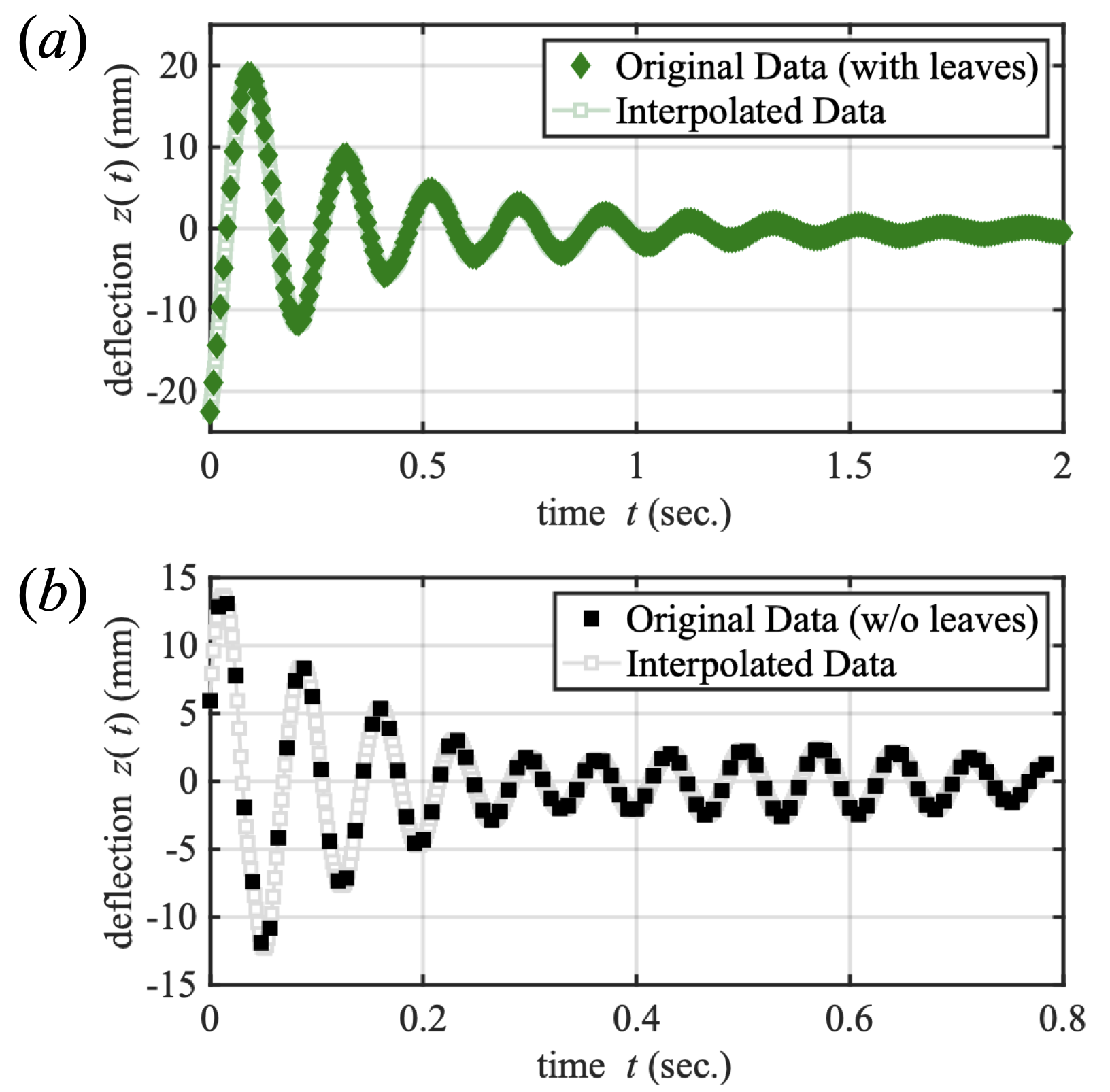}
    \caption{Waveforms obtained through analyzing the high-speed video of vibrating (a) stem with branches and leaves, and (b) stem without branches and leaves. Original data (taken at 125~f.p.s.) are marked by darker colors, while the interpolated data with spline method are marked by lighter colors. Time $t=0$ is set when a finger tip left clearly from the stem surface. Approximately first 10 cycles in vibration are presented.}
    \label{fig:vibration}
\end{figure}

We filmed the stem dynamics using a Photron Fastcam Nova 6 high-speed camera at 125~f.p.s.. The images were analyzed using Tracker software~\cite{Brown2009}, where we enabled the auto-tracking function. Unlike in the static test, we kept track of the tip of the plant whenever possible. If tracking the tip is not feasible or appears not appropriate, we tracked a visible node near the tip. Figure~\ref{fig:vibration} shows typical tracked results of the soybean stem tip displacement, $z$($t$). Figure~\ref{fig:vibration}(a) represents data for a stem with leaves and branches (green data), while Figure~\ref{fig:vibration}(b) shows results for a stem without leaves (black data). In both cases, the sway was applied at $L_\mathrm{net}\approx5$~cm. Filled markers indicate the original data taken at 125~f.p.s.. These appear relatively sparse due to the limited temporal resolution. In the analysis, we interpolated the data using a spline-type method and increased the effect sampling rate by a factor of 5, where the results are marked as shaded markers.

In our analysis, we used Matlab to detect peaks in the interpolated data. When the peak detection was successful, we adopted the logarithmic decrement method to estimate the damping ratio $\zeta$ of this (under)damped oscillation as
\begin{equation}
\zeta=\frac{\delta}{\sqrt{(2 \pi)^2+\delta^2}},
\end{equation}
where $\delta$ is the logarithmic decrement estimated as
\begin{equation}
    \delta=\frac{1}{m}\ln{\frac{z_1}{z_{m+1}}}.
\end{equation}
Here, $m$ is an integer that is chosen arbitrarily. In our analysis, we used $m=4$ by default. In cases with high damping, smaller values of $m$ were used (see details in Appendix). 
Once we know the $\zeta$ value, we may estimate the fundamental frequency of the system as
\begin{equation}
    \omega_0=\frac{\omega_d}{\sqrt{1-\zeta^2}}, ~\omega_d=m\frac{2\pi}{t_{m+1}-t_1},
\end{equation}
where $f_0=\omega_0/(2\pi)$. For the (under)damped oscillation, it is often the case that $\zeta\ll1$ and thus $\omega_0\approx\omega_d$. We note that, in some cases, the data contained noise and the peaks were not detected correctly. In such cases, we instead applied a fast Fourier transform (FFT) to estimate the value of $f_0$ (see details in Appendix).  

\subsection{Numerical Simulations}
We estimated the natural frequencies based on the distribution of leaf mass and the elastic modulus of the stems for various soybean plants, and performed finite element analysis to aid in the interpretation of the experimental results. Numerical simulations were conducted to estimate the theoretical vibration modes, excluding the influence of air. Previous research has performed similar comparisons for maize plants, based on the field experiments with the finite element simulations \cite{Nakashima2023}, which showed that the natural frequencies obtained from field experiments are almost identical to the ones from the numerical simulations.

We primarily performed the model analysis using finite element analysis (FEA) to investigate the eigenmodes of the soybean stem $\#$4 for model validation.
Table \ref{tab:femesh-geometry-params} shows the length and weight of the soybean stem $\#$4.

\begin{table}[hbtp]
  \caption{Length, diameter, and weight of the soybean model. The internode IDs were assigned starting with the fixed end as No. 1, followed by Nos. 2 and 3, and so on, up to the tip.}
  \label{tab:femesh-geometry-params}
  \centering
  \begin{tabular}{lrrrrrrrrrr}
    \hline
    Internode ID & 1 & 2 & 3 & 4 & 5 & 6  \\
    \hline
    Length (mm) & 40.0 & 40.0 & 50.0 & 95.0 & 45.0 & 30.0  \\
    Diameter (mm) & 4.5 &   4.5 &   4.0 &   3.5 &   3.0 &   2.0 \\
    Internode weight (g) & 0.142 & 0.142 &  0.188 &  0.462  &   0.213 &   0.099  \\
    Leaf weight (g) & 0.235 & 0.241 &  0.608 &  0.563  &   0.425 &   --  \\
    \hline
  \end{tabular}
\end{table}

\begin{figure}
    \centering
    \includegraphics[width=0.5\textwidth]{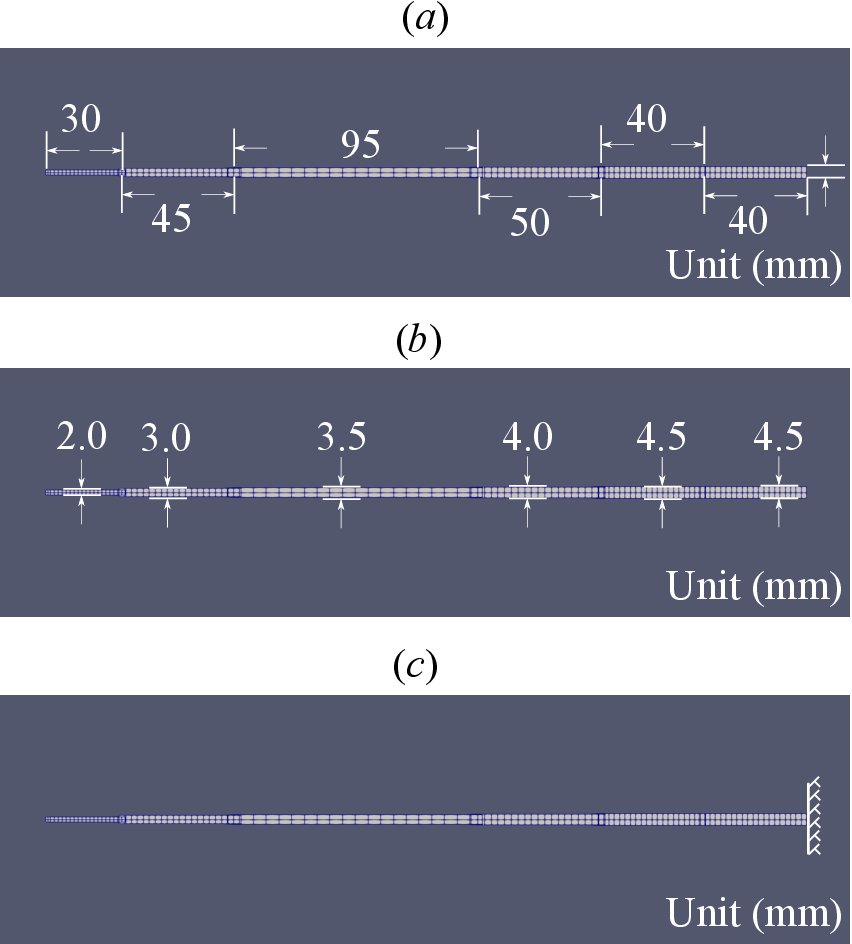}
    \caption{Numerical setup for the finite element (FE) simulations. (a, b) Initial FE mesh illustrating the length and diameter of each internode, respectively, for both the with- and without-leaves cases. (c) The fixed boundary condition applied at the bottom of the stem. The same mesh was utilized for both cases, with leaf mass incorporated directly into the corresponding internodes.}
    \label{fig:N1}
\end{figure}

Figure~\ref{fig:N1} A and B show the finite element mesh and boundary conditions for the numerical simulations. Both meshes were created based on the morphology of sample 4 (see Table~\ref{tab:femesh-geometry-params}). The Young's modulus of the stem is set to 280 MPa, which is determined from the static bending tests, and the Poisson's ratio is set to 0.30 for stems. 

In the parametric study, the Young's modulus, stem length and stem diameter listed in Table \ref{tab:femesh-geometry-params} were kept constant, and only the mass ratio was varied. The simulations were conducted by applying mass fluctuations to each section, generated from uniform random number ranging from -1 g to 1 g. Note that when the mass became negative, it was corrected by taking its absolute value. For the 456 virtual stems randomly generated using the methods described above, the natural frequencies and natural vibration modes were determined using finite element modal analysis \cite{Nakashima2023}. We used the locally optimal block preconditioned conjugate gradient (LOBPCG) method \cite{Knyazev2007} for the eigenvalue problem solver. It should be noted that the mode shape of the fundamental mode was the usual first bending mode shape in all cases, with only the natural frequency differing from case to case. Therefore, in the following, we focused solely on the natural frequencies to analyze the numerical results.

\section{Results and discussion}
\subsection{Preliminary measurement: a polycarbonate beam}

\begin{figure}
\centering
\includegraphics[width=0.95\columnwidth]{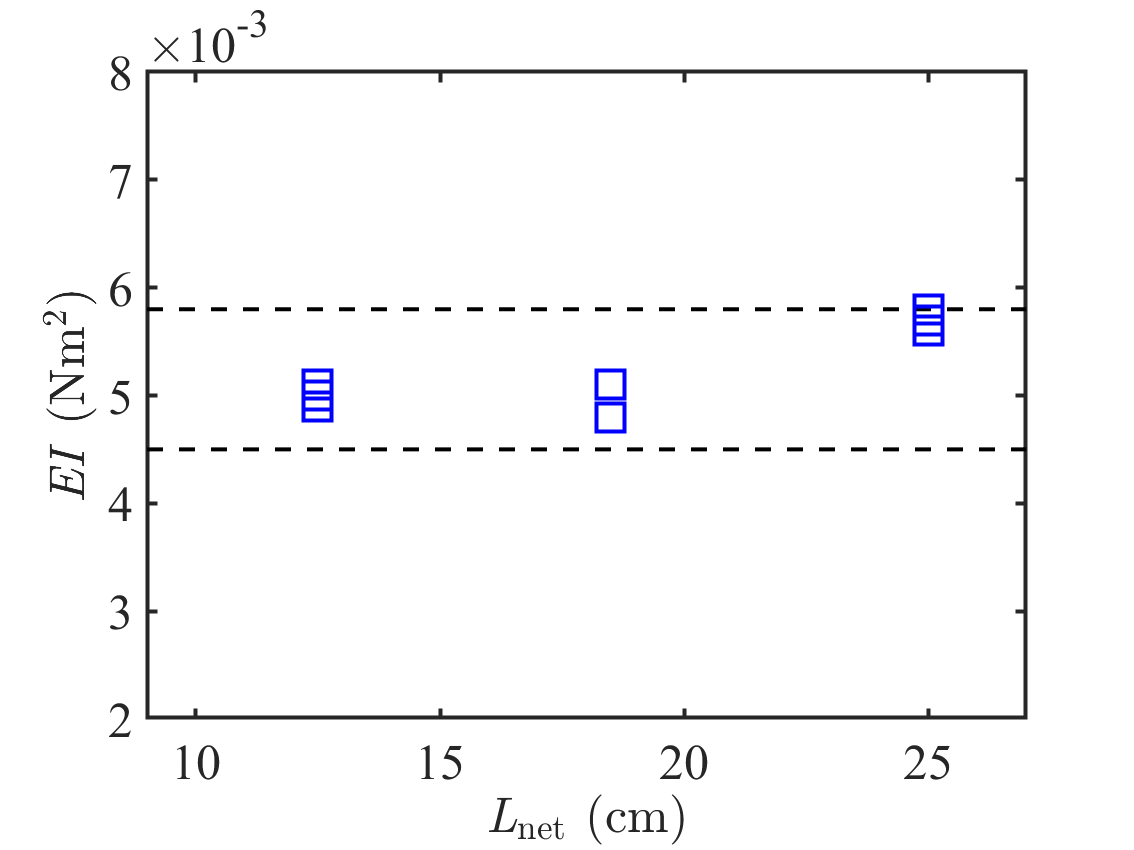}
    \caption{Flexural rigidity $EI$ of a polycarbonate beam (squares) as a function of the deflection location from the clamp, $L_\mathrm{net}$, measured via static tests. The dashed line indicates the approximate range of $EI$ estimated from the dynamic tests.}
    \label{fig:poly}
\end{figure}

We first performed both the static and the dynamic tests on a cantilever beam made of polycarbonate, where the total length of the beam $L$ from the clamp was approximately $25$~cm. We varied the location of the applied deflection $L_\mathrm{net}$ at three different lengths (12.5, 18.5, and 25~cm) and each condition was tested at least twice. The experimental data shown in Figure~\ref{fig:poly} suggest that the choice of $L_\mathrm{net}$ does not affect the result much. 

We found the averaged $EI$ value of $EI=0.00525\pm0.00039$~Nm$^2$. It was consistent with the data from the dynamic tests. The typical angular frequency ranged $\omega\approx22-25$~rad/s, meaning that $EI$ values to be $0.0044\leq EI\leq0.0058$~Nm$^2$ as marked by dashed lines as in Figure~\ref{fig:poly}. The reasonable agreement between the two approaches, within the same order of magnitude, suggests that the simple cantilever method provides a reliable estimate of the mechanical response of a homogeneous beam. 

To estimate the Young's modulus $E$, we evaluated the second moment of area $I$ to be $I=bh^3/12\approx1.22\times10^{-12}$~m$^4$, where $b\approx20$~mm and $h\approx0.9$~mm. The resulting $E$ value was found to be $E\approx4.32\pm0.32$~GPa. 
Poly-carbonate is widely used thermoplastic material. Its reported value of typical Young's modulus varies across the sources, for example, $E\approx2 - 2.44$~GPa in \cite{Ashby2005} or $E\approx1.8-3.2$~GPa as summarized in \cite{Guzzi2023}. Although a precise value for our particular sample is not available, our estimated values remain in the same order of magnitude. 
We note that our estimated value $E\approx$4.32~GPa is roughly 40\% above the value range from literature. Since both dynamic and static tests match (figure \ref{fig:poly}), this discrepancy is possibly due to the methodology. While the exact reason behind is yet unclear, it might be related to not only the clamping conditions but also some more complex factors including the three-dimensional effect~\cite{Banks1974,McShane2006} that our rather simple method does not take into account.  
Still, this simple method can give us a good estimate of the mechanical properties of the material.

\subsection{Main Experiment: Mechanical property of the soybean stem}
\subsubsection{Bending rigidity $EI$ of the soybean plant}

\begin{figure}
    \centering
\includegraphics[width=0.95\columnwidth]{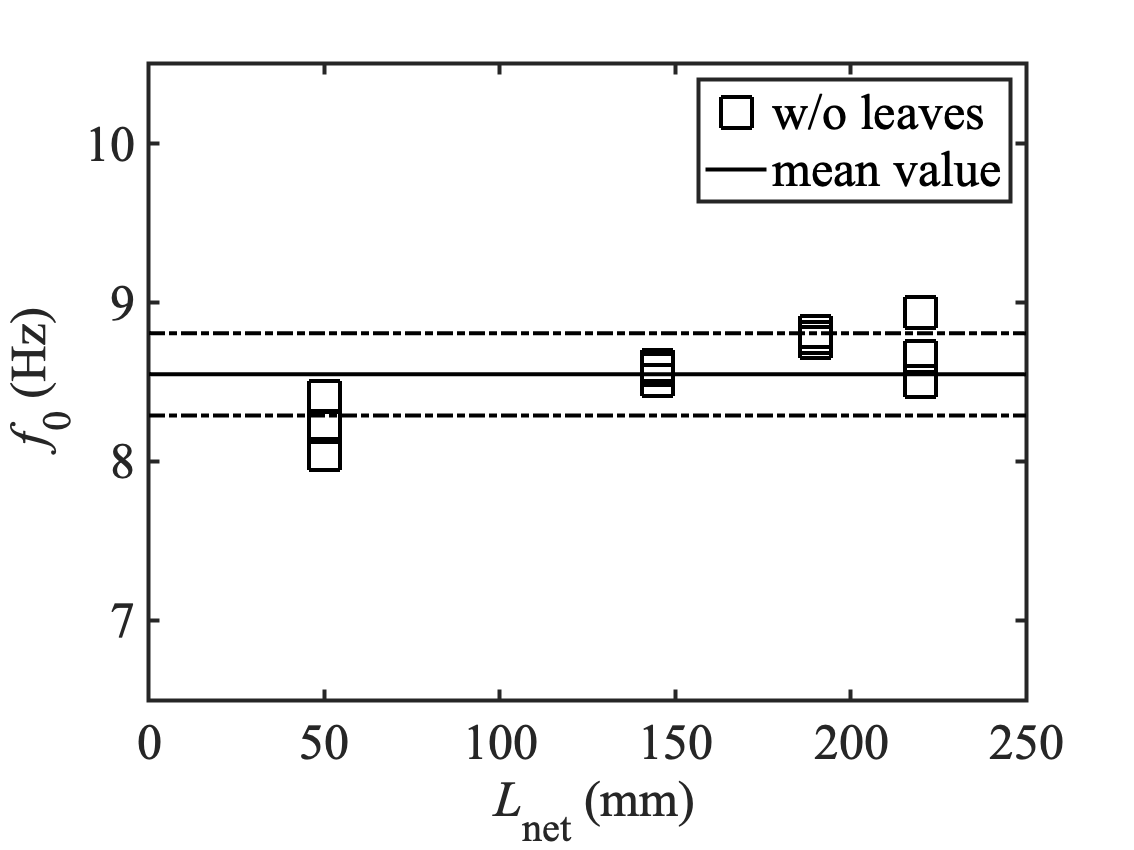}
    \caption{Frequency response of plant sample~\#4 without branches and leaves. The measured fundamental frequency is $f_0 = 8.55 \pm 0.258$~Hz.}
    \label{fig:freq_plant1_stem}
\end{figure}

We first consider the simplest case, i.e., the soybean plant without side branches and leaves. In most cases, the dynamic tests resulted in periodic vibration of the plant stem as shown in Figure \ref{fig:vibration}. Figure \ref{fig:freq_plant1_stem} shows the fundamental frequency $f_0$ obtained by the dynamic tests without branches and leaves (soybean sample \#1). The averaged fundamental frequencies $f_0\approx14.1\pm0.199$~Hz (without leaves). The frequency value does not change much for various $L_\mathrm{net}$ (i.e., the location where the perturbation is applied), suggesting that the dynamic response observed in figure \ref{fig:vibration} is repeatable as long as the stem material can be considered homogeneous. This suggests that simply letting plants sway may give us a useful method to estimate their material properties.

We summarize the $EI$ values estimated from both the dynamic (Eq. \ref{dynamic}) and static (Eq. \ref{static}) tests in Table~\ref{tab:EI_stem}. As seen in samples 1 - 3, the frequency response was consistent for various sway locations, which is indicated by small standard deviations that include data from up to four different locations. We calculated the $EI$ value based on the averaged $f_0$ and Eq. \ref{dynamic}. We also show $EI$ calculated from Eq. \ref{static} and the averaged force data up to $\Delta z/L_\mathrm{net}<5\%$ for different locations, which also resulted in reasonable agreement in terms of its order of magnitude. Although the absolute $EI$ varies between samples, it supports our rather simplistic methods is useful to estimate the material property.

We note that sample \#3 in the dynamic test recorded noticeably large standard deviations when compared to its average value. Although the full explanation is yet unclear, it is perhaps partly due to the fact that the vibration of the plant for this sample tends to attenuate quickly. It is perhaps related to the plant conditions, as the sample gets weaker as the test progresses. 

\begin{figure}
    \centering
\includegraphics[width=0.95\columnwidth]{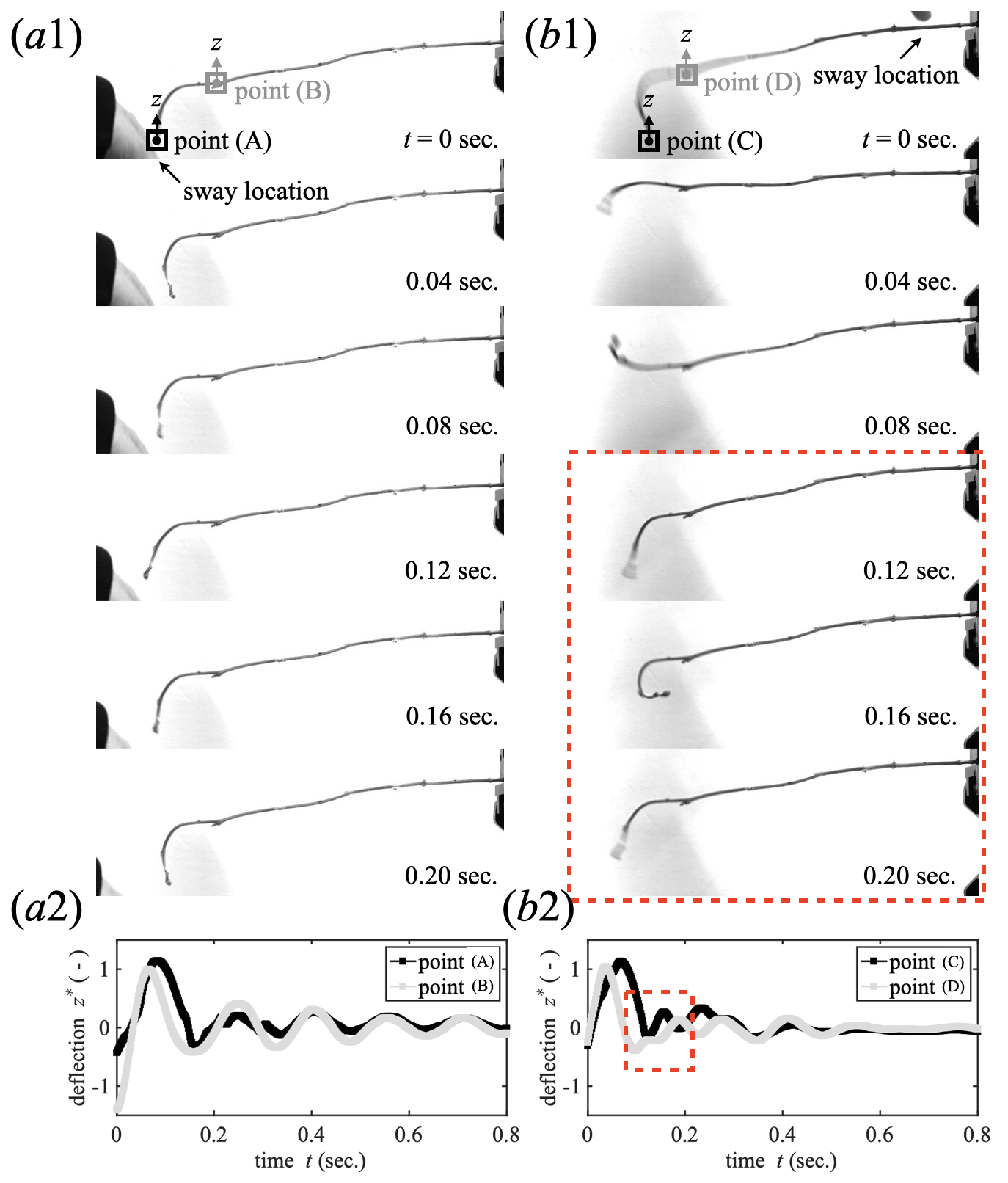}
    \caption{Displacement traces for two points on soybean sample \#3 without branches and leaves. The location of the applied sway (indicated by arrowheads) was varied between (a) $L_\mathrm{net}=29$ cm and (b) $L_\mathrm{net}=4$ cm. The tracked tip and first node are denoted by different colors. The vertical deflection, $z$, is normalized by the maximum amplitude at each respective point for direct comparison. Red squares in panels (b1) and (b2) indicate roughly synchronous time instances. }
    \label{fig:decoupledtip}
\end{figure}

Figure \ref{fig:decoupledtip} compares two cases for the plant \#3, where the sway was applied at the tip ($L_\mathrm{net}$=29~cm, see (a)) and near the clamp ($L_\mathrm{net}$=4~cm, see (b)). We marked two locations for each condition, i.e., the tip and the first node from it. The vertical axis shows the height of these two points normalized by the highest value marked by the node. Although we have used the tracked height of the tip when possible as we assumed that the stem is homogeneous, clearly, these two points move differently for the plant \#3. When we applied the sway at the tip, the vibration of the first node (point (B)) roughly corresponds to that of the tip (point (A)). On the other hand, the interaction becomes more pronounced when we applied the sway near the clamp location (see Figure~\ref{fig:decoupledtip}(b)). The tip, point (c), moves dramatically while the first node remains relatively static. Importantly, in the recovery stage of point (D) (marked by red squares), the upward motion of point (D) is clearly hindered and suppressed by the downward motion of point (C). 

As a result, the oscillation at the first node is significantly attenuated, making it difficult to reliably compute the logarithmic decrement and, consequently, the frequency. Note that it might also be related to not only the location of sway but also the time or the number of tests that the sample experienced. These data were excluded from the results listed in Table~\ref{tab:EI_stem} (see also Appendix), possibly contributing to the larger standard deviations. Note that the $EI$ values estimated from this limited dataset remain consistent with those obtained from the static tests. This suggests that the stem needs to vibrate in a synchronized manner for the dynamic tests. 


\begin{table}
    \centering
    \begin{tabular}{cccc}
        \toprule
        sample\# & $f_0$~(Hz) &$(EI)^\mathrm{dynamic}$~(Nm$^2$) & $(EI)^\mathrm{static}$~(Nm$^2$)\\
        \midrule
        1 & 14.1$\pm$0.199 & 0.0076 & 0.0038\\
        2 & 13.0$\pm$0.108 & 0.0116 & 0.0092 \\
        3 & 5.98$\pm$0.388 & 0.0044 & 0.0050 \\
        {4} & {8.55$\pm$0.258} & {0.0024} & {0.0059} \\
        \midrule
        2$^\prime$ & 13.2$\pm$0.191 & 0.0120 & 0.0095 \\
        \bottomrule
    \end{tabular}
    \caption{Frequency response of the soybean stem (without branches and leaves. $EI$ estimated by using either Eq. \ref{dynamic} or \ref{static}, corresponding to dynamic and static bending rigidity, respectively.) The sample 2$^\prime$ means the same sample but rotated $90^\circ$ to test the influence of orientation. } 
    \label{tab:EI_stem}
\end{table}

In addition, we also note that the direction of force does not significantly change the estimated $EI$ values. As summarized in Table~\ref{tab:EI_stem}, we tested plant \#2 after rotating its direction 90$^\circ$ (sample \#2$^\prime$). For the dynamic test, the fundamental frequency $f_0$ ($\approx$13.2~Hz) and the corresponding $EI$ values ($0.012$~Nm$^2$) remain very close to those of sample \#2. The static test leads to a consistent trend as well. While the number of samples is quite limited, we can conjecture that the direction is not that important for a small deformation limit. It might be practical information for on-site measurement in fields.

\subsubsection{Effects of leaves/branches}
\begin{figure}
    \centering
\includegraphics[width=0.95\columnwidth]{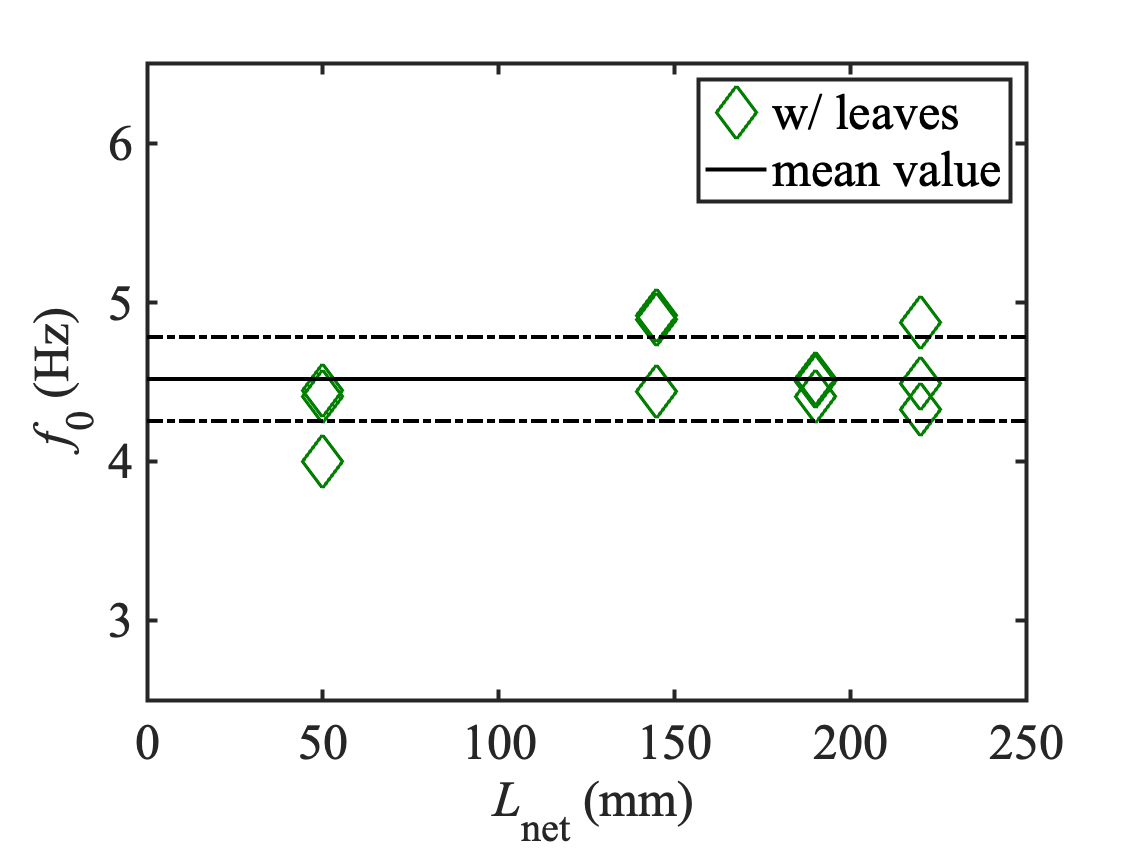}
    \caption{Frequency response of plant sample~\#4 with branches and leaves. The measured fundamental frequency is $f_0 = 4.52 \pm 0.264$~Hz.}
    \label{fig:freq_plant1_whole}
\end{figure}

Figure \ref{fig:freq_plant1_whole} shows that the fundamental frequency $f_0$ obtained by dynamic tests for plants with branches and leaves (soybean sample \#1). The fundamental frequency was $f_0\approx4.76\pm0.134$~Hz. The frequency value does not change much for various $L_\mathrm{net}$, similar to the trend reported for the plant without branches and leaves. However, it should be noted that the fundamental frequency here is significantly lower than that of the same plants without branches and leaves ($f_0\approx$14.1~Hz).

Table~\ref{tab:EI_whole} summarizes the frequency response and the $EI$ values from both dynamic and static tests. We obtained relatively small frequencies consistently. We note that the standard deviations for the plant \#2 are noticeably large. It is due to the damping for this case. In such a case, as discussed in figure \ref{fig:decoupledtip}, the logarithmic decrement routine could not be applied reliably and thus a large portion of the data was excluded (see Appendix). 

\begin{figure}
    \centering
\includegraphics[width=0.95\columnwidth]{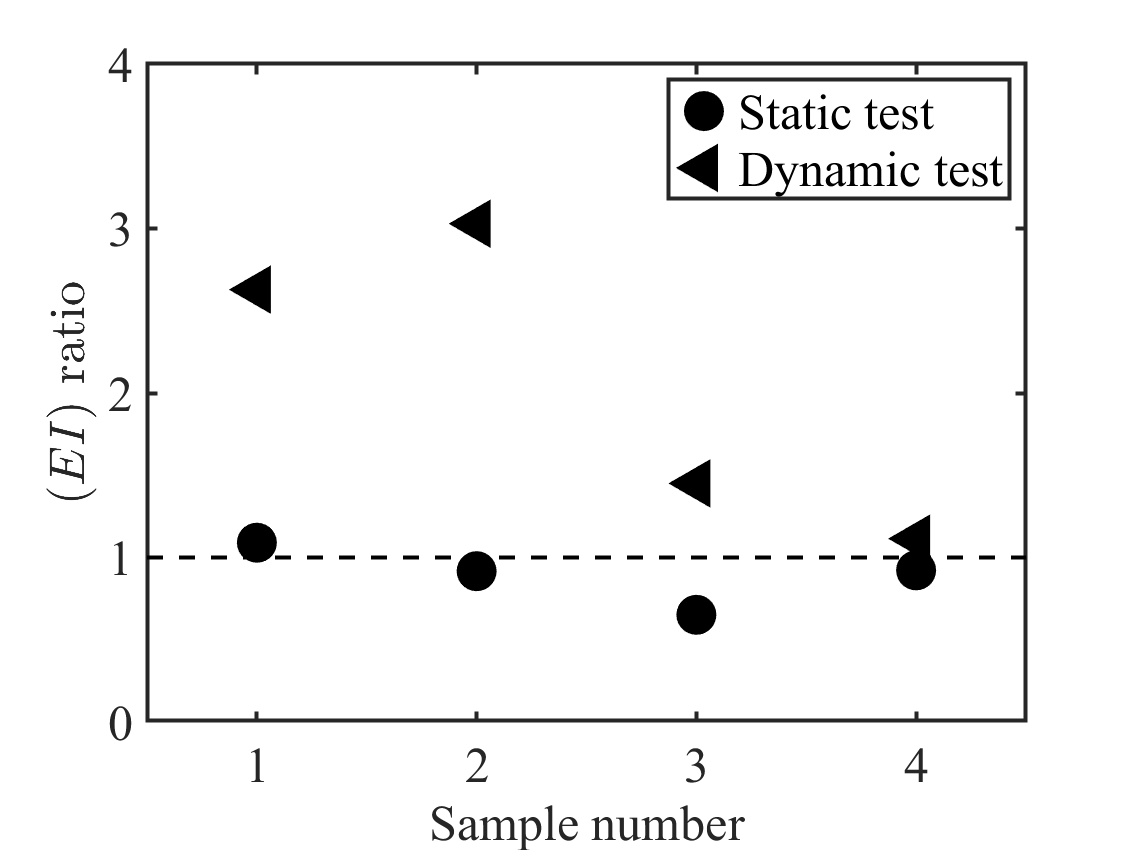}
    \caption{The ratio of $EI$ values, the estimated $EI$ values for samples without leaves and branches divided by that with leaves and branches. Marker shapes distinguishes test types. A horizontal dashed line marks the ratio of 1. }
    \label{fig:EI_ratio}
\end{figure}

Importantly, the $EI$ values obtained from the static test remain comparable to those obtained for the plant without branches and leaves (table \ref{tab:EI_stem}). 
In figure \ref{fig:EI_ratio}, we compared the ratio of $EI$ values obtained from both static (marked by circles) and dynamic (marked by triangles) tests for each sample. A horizontal dashed line represents the unity. Static test data consistently show the $EI$ ratio around unity for all samples, whereas the dynamic tests deviates largely depending on the choice of sample. Especially, samples \#1 and \#2 deviates by a factor of 3, implying the inclusion of leaves and branches affect the tendency in measurement. 
It implies that the static test with a small deflection can give us a reliable result in the $EI$ value regardless of the persistence of branches and leaves. In contrast, the $EI$ values from the dynamic tests differ noticeably. It might be worth discussing the origin of the discrepancy. In this particular sample (\#1), the soybean plant with leaves and branches shows 2.96 times lower frequency than that without them. 

The observed drop in frequency can be justified. As suggested by the relationship $f_0 \sim \sqrt{1/M}$, the addition of branches and leaves significantly increases the total mass of the system, which naturally leads to a lower natural frequency. We summarized the comparison of these ratios in Table~\ref{tab:ratio_comparison}. In all cases, $f_{\mathrm{0, whole}}$ is consistently smaller than the stem-only case ($f_{\mathrm{0, stem}}$), confirming that the added inertia of the branches and foliage plays a dominant role in the plant's dynamic response. However, the exact ratio varies between samples, suggesting that the presence of leaves does not just add simple dead weight, but likely introduces more complex dynamic interactions. This behavior may be attributed to the decoupled dynamics between the stiff stem and the more flexible branches and leaves \cite{DeLangre2019}. Because these components can vibrate out of phase with the main stem, the foliage acts as a collective mass that hinders the oscillation of the primary structure. 
The influence of the leaves discussed here might be in line with the observation in Figure~\ref{fig:decoupledtip}, where the vibration motion at the tip hinders the vibration of the nodes. The presence and relative motion of branches and leaves might result in a slightly larger damping and delay when compared to that estimated from a simple scaling based on Eq. (\ref{eq:omega}). 
This is consistent with findings in other species, such as olive plants \cite{SolaGuirado2022}, where foliage significantly alters vibration damping and causes the observed frequency shift. 

\begin{table}
    \centering
    \begin{tabular}{cccc}
        \toprule
        sample\# & $f_0$~(Hz) &$(EI)^\mathrm{dynamic}$~(Nm$^2$) & $(EI)^\mathrm{static}$~(Nm$^2$)\\
        \midrule
        1 & 4.76$\pm$0.134 & 0.0029 & 0.0034\\
        2 & 3.55$\pm$1.174 & 0.0038 & 0.0100 \\
        3 & 2.55$\pm$0.101 & 0.0031 & 0.0077 \\
        {4} & {4.52$\pm$0.264} & {0.0022} & {0.0064} \\
        \bottomrule
    \end{tabular}
    \caption{Frequency response of the soybean stem with branches and leaves. $EI$ estimated by using Eqs. 6 and 2 corresponds to those from dynamic and static tests.} 
    \label{tab:EI_whole}
\end{table}

\begin{table}
    \centering
    \begin{tabular}{ccc}
        \toprule
        sample\# & $f_\mathrm{0, whole}$/$f_\mathrm{0, stem}$ & $\sqrt{M_\mathrm{stem}/M_\mathrm{whole}}$\\
        \midrule
        1 & 0.334 & 0.546\\
        2 & 0.273 & 0.475 \\
        3 & 0.427 & 0.514 \\
        {4} & {0.530} & {0.575} \\
        \bottomrule
    \end{tabular}
    \caption{Comparison of ratios of frequency $f_\mathrm{0, whole}$/$f_\mathrm{0, stem}$ and mass $\sqrt{M_\mathrm{stem}/M_\mathrm{whole}}$.} 
    \label{tab:ratio_comparison}
\end{table}



\begin{figure}
    \centering
    \includegraphics[width=0.5\textwidth]{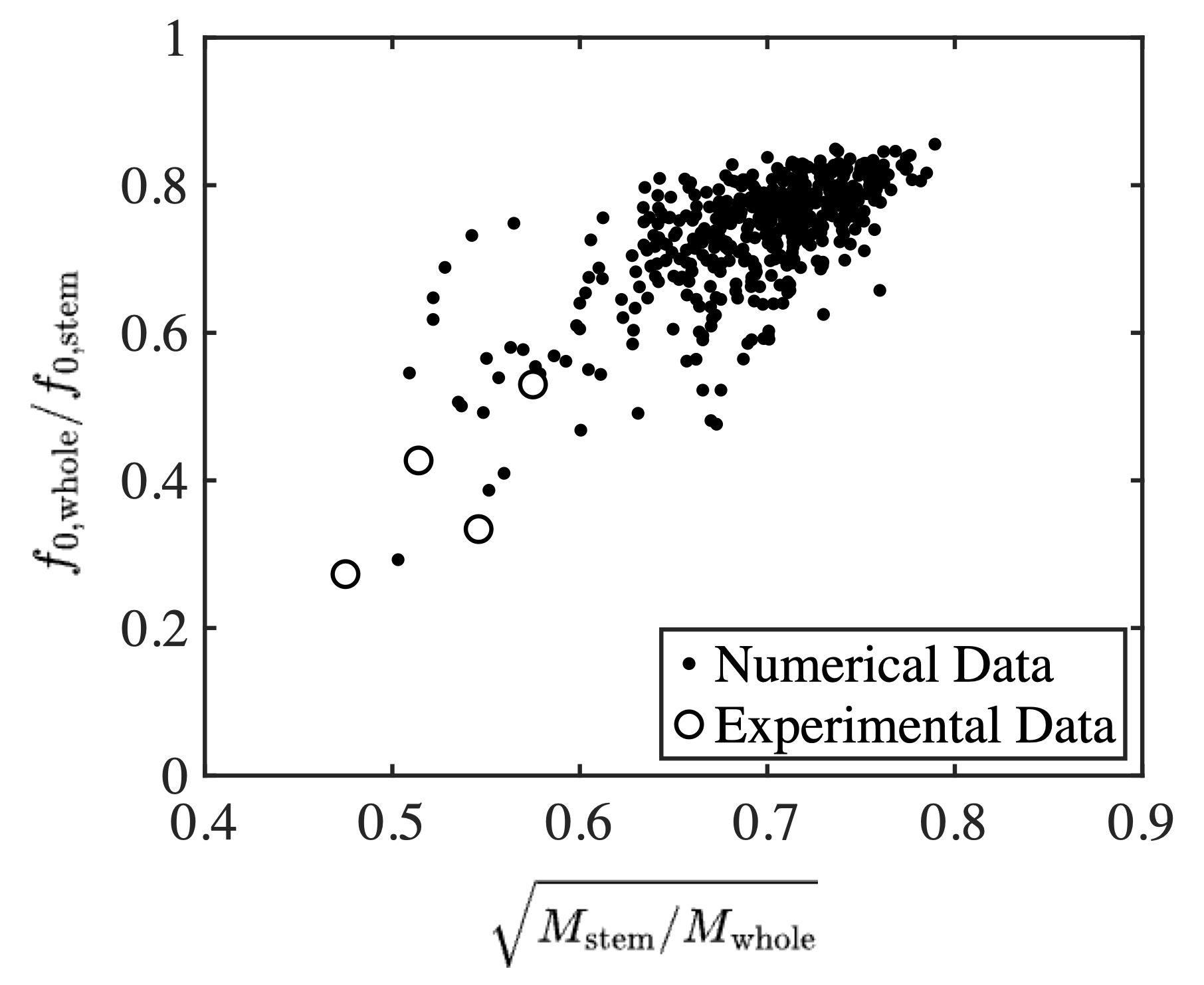}
    \caption{Parametric numerical study investigating the relationship between the frequency ratio ($f_{0,\mathrm{whole}}/f_{0,\mathrm{stem}}$) and the mass ratio ($\sqrt{M_\mathrm{stem}/M_\mathrm{whole}}$). Small filled circles denote numerical simulation results across various randomized stem structures, while open circles represent the experimental data. }
    \label{fig:fig_N4}
\end{figure}

Figure~\ref{fig:fig_N4} shows the simulation results of an investigation into the frequency-mass relationship for 1,000 cases, where randomness was introduced into the mass of the stems and leaves using uniform random numbers. The small filled circles show the numerical results, and the open circles represent experimental results. Numerical simulations suggest an incremental trend between the two dimensionless quantities with certain deviations, possibly due to randomness introduced. 
The overall trend seems to be intuitive as heavier the leaves and branches (i.e., smaller the $\sqrt{M_\mathrm{stem}/M_\mathrm{whole}}$), larger the discrepancy from the unity. The data seemingly converges to it as the relative mass of the leaves and branches becomes minor (i.e., larger $\sqrt{M_\mathrm{stem}/M_\mathrm{whole}}$ values, up to $\sqrt{M_\mathrm{stem}/M_\mathrm{whole}}\rightarrow1$). 
Experimental results agrees with a trend of numerical data for a wider range of random plant structures. It implies that the additional mass has certain level of impact on the dynamic response, as expected.

\section{Conclusion and Discussions}
This paper evaluates experimental methods to estimate the mechanical property, i.e., the rigidity of the plant. We employed both the dynamic vibration and static force tests, where both showed good agreement for a homogeneous material (Figure~\ref{fig:poly}). We used a soybean plant for the experimental sample and tested it with and without branches and leaves. The location of sway/force applications, $L_\mathrm{net}$, was varied.

Dynamic vibration test consistently resulted in robust $EI$ values, when compared to the static test, in sensitive to either the choice of $L_\mathrm{net}$ (figure \ref{fig:EI_ratio}) or the direction of the stem (table \ref{tab:EI_stem}).  
It implies that the dynamic force tests worked well when branches and leaves are removed and as long as the stem vibrates uniformly, i.e., when the decoupled motion (figure \ref{fig:decoupledtip}) is not present. However, when the branches and leaves remain attached to the stem, the dynamic response may change dramatically, possibly due to the additional mass $M$ (see table~\ref{tab:ratio_comparison} and figure~\ref{fig:fig_N4}) and perhaps the dynamic interaction between them (see figure~\ref{fig:decoupledtip}). The involvement of these factors may limit the applicability of the technique to the on-site measurement, as the mass (either $M_\mathrm{stem}$ or $M_\mathrm{whole}$) and detailed structural information of plant are usually not accessible in the field.
It implies that rather simple cantilever force measurement is quite useful for estimating the $EI$, which is insensitive to the presence of branches  and leaves as suggested in figure~\ref{fig:EI_ratio}. Even in field conditions, as this method essentially requires only the force $F$, deflection $\Delta z$, and the length $L_\mathrm{net}$, adaptation would be relatively straightforward. 

\begin{figure}
    \centering
\includegraphics[width=0.95\columnwidth]{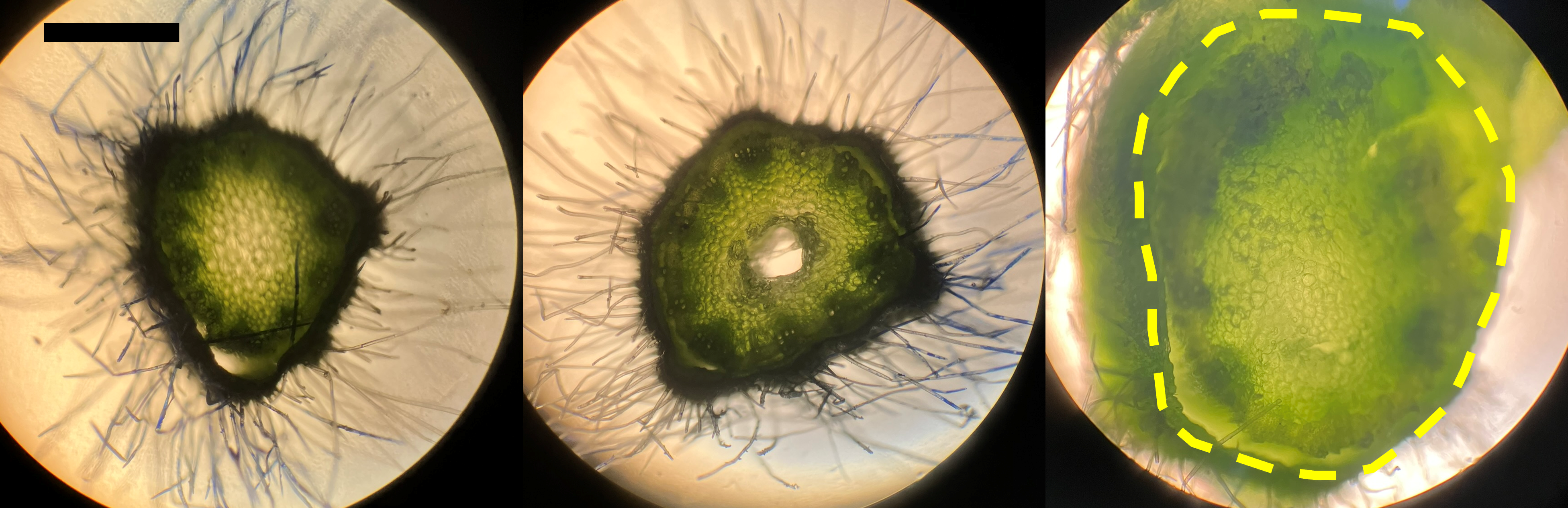}
    \caption{ Cross-sectional images of plant \tb{\#2} at different locations ($L_\mathrm{net}$). The left, middle, and right panels correspond to $L_\mathrm{net}\simeq20$~cm (3~cm from the tip), $L_\mathrm{net}\simeq16$~cm (7~cm from the tip), and $L_\mathrm{net}\simeq10$~cm (13~cm from the tip), respectively. For some samples, the large slice size caused overlap; therefore, the edges are outlined with yellow dashed lines for clarity. Scale bars represent 1~mm. }
    \label{fig:crosssection}
\end{figure}

Lastly, but not least, there are a few caveats for on-site measurements. While overall rigidity to the bending can be captured through the $EI$ value, the Young's modulus $E$ would be an important parameter that describes the material property. The second moment of area $I$ is essential to compute the Young's modulus $E$. However, the outer radius $R_\mathrm{out}$ differs along with the stem axis, implying the necessity of repeated measurements at different locations. 
As an example, we cut the stem after experimental runs to measure its size and to observe its structure (figure \ref{fig:crosssection}). It is obvious that the outer radius, which is the most convenient length to measure, varies along the stem length. We measured at different locations for each stem, resulting in obtaining factors ranging from 1.16 to 1.67, which is monotonically neither increases nor decreases. This deviation is must be significant as the second moment of area for a cylinder can be expressed as $I\sim R^4$. 
Moreover, another major source of uncertainty in estimating $E$ is due to the hollow structure of the stem. 
As visualized in figure \ref{fig:crosssection}, the stem contains both a relatively stiff outer region and a softer or even open inner structure. As the first-order consideration, we calculated $I_\mathrm{open}/I_\mathrm{filled}$ when there are hollow structures observed.
The second moment of area $I_\mathrm{open}$ is calculated by $I_\mathrm{open}=\pi(R^4_\mathrm{out}-R^4_\mathrm{in})/4$, where $R_\mathrm{out}$ and $R_\mathrm{in}$ are evaluated at each $L_\mathrm{net}$. $I_\mathrm{filled}$ can be estimated by letting $R_\mathrm{in}=0$. 
It implies that, though its influence might be limited when compared to that of variation in radius along the stem axis, the cross-sectional structure can possibly affect the resulting $E$ values even though the materials are sufficiently homogeneous. Note that, the inner structure could be influenced by not only the size of the stem but also the water-stressed condition and not observable from outside. The potential contribution of $I_\mathrm{open}/I_\mathrm{filled}$ needs to be considered as an additional caveat when performing non-invasive measurements on-site, where the inner structure data are perhaps unavailable.

It is perhaps worth discussing the estimation of Young’s modulus $E$ from our $EI$ measurement, despite the limitations noted above. We estimated $EI$ values based on the static force measurement in a small deflection limit. 
Table~\ref{tab:E_stem} summarizes the Young's modulus $E$ estimated based on the force measurement. 
The data remain similar in magnitude regardless of the presence of branches and leaves, as expected. Within the limited accuracy of our approach, the results showed that the Young's modulus $E$ of the tested plants lies in the range between $E\sim {\mathcal O}(10) - {\mathcal O}(100)$~MPa. It is apparent that our estimated $E$ values differ from the reported value (i.e., elastic modulus of $\sim{\mathcal{O}}(10)$~GPa) for a dried soybean stalk \cite{Wu2010} and other species summarized in the Introduction. It could be attributed to the hydration level. In a recent report \cite{Xu2020}, they employed three-point bending method and reported that the breaking force of the stem increases as the growth period of the plant progresses, while the maximum deformation decreases. While there is no explicit conversion to the $E$ value in their report, it is implied that the growth stage and its hydration have certain impact on the net strength and material properties of the stem. Not only the one discussed above but also others including growth environment would be important directions for follow-up studies. 

\begin{table}
    \centering
    \begin{tabular}{l|p{35mm}|p{35mm}}
        \toprule
        sample\# & $E$ (MPa) \newline w/ branches \& leaves & $E$ (MPa) \newline w/o branches \& leaves \\
        \midrule
        1 & 8.83 - 61.9 & 10.3 - 65.5\\
        2 & 27.8 - 334 & 28.1 - 274\\
        3 & 128 - 583 & 159 - 229\\
        \midrule
        2$^\prime$ & N/A &26.9 - 290 \\
        \bottomrule
    \end{tabular}
    \caption{Estimated Young's modulus $E$ based on the force measurement (static test). The second moment of area $I$ is estimated based on the inner and outer radii $R_\mathrm{in}$ and $R_\mathrm{out}$ of the stem at each $L_\mathrm{net}$. The range corresponds to the maximum and the minimum values over measurements in different $L_\mathrm{net}$. Sample \#2$^\prime$ represents the rotated experiment (see table \ref{tab:EI_stem}).} 
    \label{tab:E_stem}
\end{table}

\section{Data availability statement}
Data are available at OSF website https://doi.org/10.17605/OSF.IO/HX3AB .  

\section{Competing interest} The authors declare no conflict of interests

\section{Author contributions}
S.J. conceptualized the work; A.K. designed the experiments; J. Y., S.W. and A. K. conducted the experiments and analyzed data; H. T. performed numerical simulation; S. J. supervised research; A.K. wrote the original draft of the manuscript, and all other authors discussed the results and edited the manuscript.

\section{Funding statement} 

J.Y, S.W, and S.J. were supported by the National Science Foundation CBET-2401507. H.T were supported by JSPS KAKENHI Grant Numbers JP24K01739, JP23K18019 and JP20K22599.

\section{Use of AI tools}
The authors used the AI tools to correct grammatical errors in the manuscript. 

\section*{Appendix: List of analyzed data in dynamic tests}
We summarized the data subjected to the dynamic test in Table \ref{tab:FreqAnalData}. We used four different trees, which are identified through sample number and the location of sway $L_\mathrm{net}$ (see figure \ref{fig:setup}). We repeated the tests three times at each condition, while some data were either not available or able to be analyzed. As stated in Section II.B, we employed logarithmic decrement $\delta$ to estimate damping ratio $\zeta$ as well as the fundamental frequency $f_0$. We set $m=4$ by default to calculate $\delta$ (see equation (6)). However, in some cases, we resorted to use smaller $m$ values. We also used a fast Fourier transform when the method above was not applicable. While we have some data missing in the table \ref{tab:FreqAnalData}, we could attain sufficient number of data to capture the general trend; note that the difference in number of data that could be analyzed in dynamic test might contribute to the variation of $f_0$ (see tables \ref{tab:EI_stem} \& \ref{tab:EI_whole}).

\begin{table*}[h]
    \centering
    \caption{List of data subjected to the dynamic test. Sample number, location of sway ($L_\mathrm{net}$), number of trials subjected to the analysis, particular method(s) adopted, damping ratio $\zeta$, and the estimated fundamental frequency $f_0$, were respectively summarized.}
    \label{tab:FreqAnalData}
   { \begin{tabular}{|l|l|l|l|l|l|}
        \hline \hline
         Sample \# & $L_\mathrm{net} (mm) $ & \# of trials analyzed & Methods & $\zeta$ &$f_0$ (Hz)  \\ \hline

        \multirow{4}{*}{1 (w/ leaves)} & 5 & 3 & Log. Decrement ($m=4$) \& FFT & 0.095 &4.7--5.0 \\ 
        {} & 10 & 3 & Log. Decrement ($m=2$) \& FFT & 0.15 &4.6 \\ 
        {} & 17 & 3  & Log. Decrement ($m=2$) \& FFT & 0.14 &4.7--4.8 \\ 
        {} & 21 & 3 & Log. Decrement ($m=4$) & 0.089--0.10 &4.8--4.9\\ \hline
        
        \multirow{4}{*}{1 (w/o leaves)} & 5 & 3 & Log. Decrement ($m=4$) & 0.057--0.082 &13.9--14.1 \\ 
        {} & 10 & 3 & Log. Decrement ($m=4$) & 0.062--0.11 &13.8--14.5 \\ 
        {} & 17 & 3  & Log. Decrement ($m=1,4$) & 0.060--0.14 &12.4--14.1 \\ 
        {} & 21 & 3 & Log. Decrement ($m=4$) & 0.051--0.083 &14.2--14.4\\ \hline\hline

         \multirow{4}{*}{2 (w/ leaves)} & 10 & 1 & Log. Decrement ($m=4$) & 0.69 &5.3 \\ 
        {} & 16 & 1 & Log. Decrement ($m=4$) & 0.38 &2.8 \\ 
        {} & 20 & 1 & Log. Decrement ($m=4$) & 0.29 &3.0 \\ 
        {} & 23 & 1 & Log. Decrement ($m=4$) & 0.17 &3.0 \\ \hline
        
        \multirow{4}{*}{2 (w/o leaves)} & 10 & 3 & Log. Decrement ($m=4$) & 0.040--0.044 &12.8--12.9 \\ 
        {} & 16 & 3 & Log. Decrement ($m=4$) & 0.032--0.039 &13.0 \\ 
        {} & 20 & 3 & Log. Decrement ($m=4$) & 0.031--0.035 &13.0--13.1 \\ 
        {} & 23 & 3 & Log. Decrement ($m=4$) & 0.030--0.040 &13.0--13.2 \\ \hline

        \multirow{4}{*}{2 (w/o leaves, rotated)} & 10 & 3 & Log. Decrement ($m=4$) & 0.054--0.065 &12.9--13.1 \\ 
        {} & 16 & 3 & Log. Decrement ($m=4$) & 0.057--0.086 &13.1 \\ 
        {} & 20 & 3 & Log. Decrement ($m=4$) & 0.045--0.064 &13.2--13.5 \\ 
        {} & 23 & 3 & Log. Decrement ($m=4$) & 0.029--0.039 &13.4--13.5 \\ \hline\hline

       \multirow{4}{*}{3 (w/ leaves)} & 4 & 2 & Log. Decrement ($m=1$) & 0.19--0.24 &2.4 \\ 
        {} & 14 & 1 & Log. Decrement ($m=1$) & 0.26 &2.5 \\ 
        {} & 23 & 3 & Log. Decrement ($m=4$) & 0.097--0.11 & 2.6 \\ 
        {} & 29 & 2 & Log. Decrement ($m=1,4$) & 0.067--0.13 &2.6 \\ \hline

        \multirow{4}{*}{3 (w/o leaves)} & 4 & 0 & N/A & N/A & N/A \\ 
        {} & 14 & 0 & N/A & N/A & N/A \\ 
        {} & 23 & 2 & Log. Decrement ($m=1$) & 0.15--0.22 & 6.3--6.5 \\ 
        {} & 29 & 3 & Log. Decrement ($m=1$) & 0.010--0.014 &5.5--5.9 \\ \hline\hline
        
        \multirow{4}{*}{4 (w/ leaves)} & 5 & 2 & Log. Decrement ($m=4$) \& FFT & 0.18 &4.0--4.4 \\ 
        {} & 14.5 & 3 & Log. Decrement ($m=4$) & 0.11--0.18 &4.4--4.9 \\ 
        {} & 19 & 3 & Log. Decrement ($m=4$) & 0.16--0.20 & 4.4--4.5 \\ 
        {} & 22 & 0 & N/A & N/A & N/A \\ \hline
        
        \multirow{4}{*}{4 (w/o leaves)} & 5 & 3 & Log. Decrement ($m=4$) & 0.074--0.11 &8.0--8.4 \\ 
        {} & 14.5 & 3 & Log. Decrement ($m=4$) & 0.067--0.097 &8.5--8.6 \\ 
        {} & 19 & 3 & Log. Decrement ($m=4$) & 0.031--0.061 & 8.7--8.8 \\ 
        {} & 22 & 3 & Log. Decrement ($m=4$) & 0.014--0.044 &8.5--8.9 \\ \hline
    \end{tabular}}
\end{table*}

\bibliography{Haruka}

@PREAMBLE{
 "\providecommand{\noopsort}[1]{}" 
 # "\providecommand{\singleletter}[1]{#1}%" 
}

@article{Berry2004,
author = {Berry, P M and Sterling, M and Spink, J H and Baker, C J and Sylvester-Bradley, R. and Mooney, S J and Tams, A R and Ennos, A R},
journal = {Advances in Agronomy},
number = {4},
pages = {217--271},
title = {{Understanding and reducing lodging in cereals}},
volume = {84},
year = {2004}
}

@article{Baker1995,
author = {Baker, C. J.},
doi = {10.1006/jtbi.1995.0147},
issn = {00225193},
journal = {Journal of Theoretical Biology},
number = {3},
pages = {355--372},
title = {{The development of a theoretical model for the windthrow of plants}},
volume = {175},
year = {1995}
}

@article{Berry2003,
author = {Berry, P. M. and Spink, J. and Sterling, M. and Pickett, A. A.},
journal = {Journal of Agronomy and Crop Science},
number = {6},
pages = {390--401},
title = {{Methods for rapidly measuring the lodging resistance of wheat cultivars}},
volume = {189},
year = {2003}
}

@article{Shu2020,
author = {Shu, Meiyan and Zhou, Longfei and Gu, Xiaohe and Ma, Yuntao and Sun, Qian and Yang, Guijun and Zhou, Chengquan},
doi = {10.1016/j.asr.2019.09.034},
isbn = {2448858578},
issn = {18791948},
journal = {Advances in Space Research},
number = {1},
pages = {470--480},
title = {{Monitoring of maize lodging using multi-temporal Sentinel-1 SAR data}},
volume = {65},
year = {2020}
}

@article{Huang2016,
author = {Huang, Jiale and Liu, Wangyu and Zhou, Feng and Peng, Yujian and Wang, Ningling},
doi = {10.1016/j.biosystemseng.2016.09.016},
issn = {15375110},
journal = {Biosystems Engineering},
pages = {298--307},
publisher = {Elsevier Ltd},
title = {{Mechanical properties of maize fibre bundles and their contribution to lodging resistance}},
url = {http://dx.doi.org/10.1016/j.biosystemseng.2016.09.016},
volume = {151},
year = {2016}
}

@article{Oduntan2024,
author = {Oduntan, Yusuf and Kunduru, Bharath and Tabaracci, Kaitlin and Mengistie, Endalkachew and McDonald, Armando G. and Sekhon, Rajandeep S. and Robertson, Daniel J.},
doi = {10.1016/j.eja.2024.127262},
issn = {11610301},
journal = {European Journal of Agronomy},
number = {February},
pages = {127262},
publisher = {Elsevier B.V.},
title = {{The effect of structural bending properties versus material bending properties on maize stalk lodging}},
url = {https://doi.org/10.1016/j.eja.2024.127262},
volume = {159},
year = {2024}
}

@article{Ottesen2023,
author = {Ottesen, Michael and Carter, Joseph and Hall, Ryan and Liu, Nan Wei and Cook, Douglas D.},
doi = {10.1093/insilicoplants/diad010},
issn = {25175025},
journal = {In Silico Plants},
number = {2},
pages = {1--13},
title = {{Development and stochastic validation of a parameterized model of maize stalk flexure and buckling}},
volume = {5},
year = {2023}
}

@article{VonForell2015,
author = {{Von Forell}, Greg and Robertson, Daniel and Lee, Shien Yang and Cook, Douglas D.},
doi = {10.1093/jxb/erv108},
issn = {14602431},
journal = {Journal of Experimental Botany},
number = {14},
pages = {4367--4371},
pmid = {25873674},
title = {{Preventing lodging in bioenergy crops: A biomechanical analysis of maize stalks suggests a new approach}},
volume = {66},
year = {2015}
}

@article{Berry2021,
author = {Berry, P. M. and Baker, C. J. and Hatley, D. and Dong, R. and Wang, X. and Blackburn, G. A. and Miao, Y. and Sterling, M. and Whyatt, J. D.},
doi = {10.1016/j.fcr.2020.108037},
issn = {03784290},
journal = {Field Crops Research},
number = {July 2020},
pages = {108037},
publisher = {Elsevier B.V.},
title = {{Development and application of a model for calculating the risk of stem and root lodging in maize}},
url = {https://doi.org/10.1016/j.fcr.2020.108037},
volume = {262},
year = {2021}
}

@article{Ogilvie2025,
author = {Ogilvie, Grant and Cook, Douglas},
doi = {10.1109/ietc64455.2025.11039488},
isbn = {9798331512828},
journal = {2025 Intermountain Engineering, Technology and Computing (IETC)},
pages = {1--6},
publisher = {IEEE},
title = {{Measuring the Effects of the Leaf Blades and Leaf Sheaths on the Dynamic Behavior of Maize}},
year = {2025}
}

@article{AlZube2018,
   author = {Loay, Al‑Zube and Wenhuan, Sun and Daniel, Robertson
and Douglas, Cook},
   title = {The elastic modulus for maize stems},
   journal = {Plant Methods},
   year = {2018},
   volume = {14},
   issue = {11},
   pages = {1--12},
}

@article{Nakashima2023,
   author = {Taiken Nakashima and Haruka Tomobe and Takumi Morigaki and Mengfan Yang and Hiroto Yamaguchi and Yoichiro Kato and Wei Guo and Vikas Sharma and Harusato Kimura and Hitoshi Morikawa},
   title = {Non‑destructive high‑throughput measurement of elastic‑viscous properties of maize using a novel ultra‑micro sensor array and numerical validation},
   journal = {Scientifc Reports},
   year = {2023},
   volume = {13},
   issue = {},
   pages = {4914},
}

@article{Shah2016,
   author = {Darshil, U. Shah and Thomas, P.S. Reynolds and Michael, H. Ramage},
   title = {The strength of plants: theory and experimental methods to measure the mechanical properties of stems},
   journal = {Journal of Experimental Botany},
   year = {2016},
   volume = {68},
   issue = {16},
   pages = {4497–-4516},
}

@article{Caliaro2013,
   author = {Marco, Caliaro and Florian, Schmitch and Thomas, Speck and Olga, Speck},
   title = {EFFECT OF DROUGHT STRESS ON BENDING STIFFNESS IN
PETIOLES OF CALADIUM BICOLOR (ARACEAE)},
   journal = {American Journal of Botany},
   year = {2013},
   volume = {100},
   issue = {11},
   pages = {2141--2148},
}

@article{SolaGuirado2022,
title = {Effect of leaves in the dynamic response of olive tree branches and their computational model},
journal = {Computers and Electronics in Agriculture},
volume = {203},
pages = {107490},
year = {2022},
author = {Rafael R. Sola-Guirado and Rafael Luque-Mohedano and Sergio Tombesi and Gregorio Blanco-Roldan},
}

@article{DeLangre2019,
    author = {de Langre, Emmanuel},
    title = {Plant vibrations at all scales: a review},
    journal = {Journal of Experimental Botany},
    volume = {70},
    number = {14},
    pages = {3521-3531},
    year = {2019},
}

@article{Brown2009,
   author = {Douglas Brown and Anne J. Cox},
   doi = {10.1119/1.3081296},
   issn = {0031-921X},
   issue = {3},
   journal = {The Physics Teacher},
   month = {3},
   pages = {145-150},
   publisher = {American Association of Physics Teachers (AAPT)},
   title = {Innovative Uses of Video Analysis},
   volume = {47},
   year = {2009},
}

@article{Knyazev2007,
archivePrefix = {arXiv},
arxivId = {0705.2626},
author = {Knyazev, A. V. and Argentati, M. E. and Lashuk, I. and Ovtchinnikov, E. E.},
doi = {10.1137/060661624},
eprint = {0705.2626},
issn = {10648275},
journal = {SIAM Journal on Scientific Computing},
number = {5},
pages = {2224--2239},
title = {{Block Locally Optimal Preconditioned Eigenvalue Xolvers (BLOPEX) in hypre and PETSc}},
volume = {29},
year = {2007}
}

@article{Brune2018,
author = {Brune, Philip F. and Baumgarten, Andy and McKay, Steve J. and Technow, Frank and Podhiny, John J.},
doi = {10.1007/s11104-017-3457-9},
issn = {15735036},
journal = {Plant and Soil},
number = {1-2},
pages = {397--408},
publisher = {Plant and Soil},
title = {{A biomechanical model for maize root lodging}},
volume = {422},
year = {2018}
}

@article{Baker1998,
author = {Baker, C. J. and Berry, P. M. and Spink, J. H. and Sylvester-Bradley, R. and Griffin, J. M. and Scott, R. K. and Clare, R. W.},
journal = {Journal of Theoretical Biology},
number = {4},
pages = {587--603},
title = {{A method for the assessment of the risk of wheat lodging}},
volume = {194},
year = {1998}
}

@article{Niklas1988,
author = {Niklas, Karl J and Moon, Francis C},
doi = {https://doi.org/10.1002/j.1537-2197.1988.tb11225.x},
journal = {American Journal of Botany},
number = {10},
pages = {1517--1525},
title = {{Flexural stiffness and modulus of elasticity of flower stalks from allium sativum as measured by multiple resonance frequency spectra}},
url = {https://bsapubs.onlinelibrary.wiley.com/doi/abs/10.1002/j.1537-2197.1988.tb11225.x},
volume = {75},
year = {1988}
}

@BOOK{Piper1867,
   author       = {Piper , Charles Vancouver and Morse, William Joseph},
   year         = 1967,
   title        = {The soy bean: history, varieties and field studies},
   publisher    = {Washington : G.P.O.}
}

@article{Yamaguchi2014,
author = {Yamaguchi, Naoya and Sayama, Takashi and Yamazaki, Hiroyuki and Miyoshi, Tomoaki and Ishimoto, Masao},
doi = {10.1270/jsbbs.64.300},
journal = {Breeding Science},
pages = {300--308},
title = {{Quantitative trait loci associated with lodging tolerance in soybean cultivar ‘ Toyoharuka '}},
volume = {64},
year = {2014}
}

@article{Li2024,
author = {Li, Weiwei and Wang, Lei and Xue, Hong and Zhang, Mingming and Song, Huan and Qin, Meng and Dong, Quanzhong},
doi = {10.3389/fpls.2024.1477616},
issn = {1664462X},
journal = {Frontiers in Plant Science},
number = {October},
pages = {1--7},
title = {{Molecular and genetic basis of plant architecture in soybean}},
volume = {15},
year = {2024}
}

@article{Umburanas2022,
author = {Umburanas, Renan Caldas and Kawakami, Jackson and Ainsworth, Elizabeth Anna and Favarin, Jos{\'{e}} La{\'{e}}rcio and Anderle, Leonardo Zabot and Dourado-Neto, Durval and Reichardt, Klaus},
doi = {10.1038/s41598-021-04043-8},
isbn = {0123456789},
issn = {20452322},
journal = {Scientific Reports},
number = {1},
pages = {1--14},
pmid = {35017557},
publisher = {Nature Publishing Group UK},
title = {{Changes in soybean cultivars released over the past 50 years in southern Brazil}},
url = {https://doi.org/10.1038/s41598-021-04043-8},
volume = {12},
year = {2022}
}

@article{Tomobe2019,
author = {Tomobe, Haruka and Fujisawa, Kazunori and Murakami, Akira},
doi = {10.1016/j.sandf.2019.08.003},
issn = {00380806},
journal = {Soils and Foundations},
number = {6},
pages = {1860--1874},
publisher = {Japanese Geotechnical Society},
title = {{Experiments and FE-analysis of 2-D root-soil contact problems based on node-to-segment approach}},
url = {https://doi.org/10.1016/j.sandf.2019.08.003},
volume = {59},
year = {2019}
}

@article{yuk2026leaf,
  title={Leaf-inspired rain-energy harvesting device},
  author={Yuk, Jisoo and Leem, Alicia and Thomas, Kate and Jung, Sunghwan},
  journal={Physical Review Applied},
  volume={25},
  number={3},
  pages={034049},
  year={2026},
  publisher={APS}
}

@article{jung2021measuring,
  title={Measuring fluttering frequency of a leaf under water stress},
  author={Jung, Sunghwan and Yuk, Jisoo and Fuchs, Matthieu},
  year={2021},
  journal={ESS Open Archive},
  publisher={ESS Open Archive}
}

@inproceedings{yuk2022visual,
  title={Visual measurements of fluttering leaf to quantify internal water stress},
  author={Yuk, Jisoo and Lee, Joseph and Graves, Caroline and Jung, Sunghwan},
  booktitle={2022 ASABE Annual International Meeting},
  pages={1},
  year={2022},
  organization={American Society of Agricultural and Biological Engineers}
}

@inproceedings{fuchs2021fluttering,
  title={Fluttering leaves to quantify leaf's stiffness},
  author={Fuchs, Matthieu and Hooshanginejad, Alireza Navid and Yuk, Jisoo and Jung, Sunghwan},
  booktitle={2021 ASABE Annual International Virtual Meeting},
  pages={1},
  year={2021},
  organization={American Society of Agricultural and Biological Engineers}
}

@article{gart2015droplet,
  title={Droplet impacting a cantilever: A leaf-raindrop system},
  author={Gart, Sean and Mates, Joseph E and Megaridis, Constantine M and Jung, Sunghwan},
  journal={Physical Review Applied},
  volume={3},
  number={4},
  pages={044019},
  year={2015},
  publisher={APS}
}

@article{bhosale2020bending,
  title={Bending, twisting and flapping leaf upon raindrop impact},
  author={Bhosale, Yashraj and Esmaili, Ehsan and Bhar, Kinjal and Jung, Sunghwan},
  journal={Bioinspiration \& Biomimetics},
  volume={15},
  number={3},
  pages={036007},
  year={2020},
  publisher={IOP Publishing}
}

@book{Ashby2005,
    author = {Michael, F. Ashby},
    title = {Materials Selection in Mechanical Design},
    publisher = {Elsevier, Butterworth-Heinemann},
    year = {2005}
}

@book{Guzzi2023,
    author = {Francesco Guzzi and Elvira Parrotta and Simona Zaccone and Tania Limongi and Giovanni Cuda and Gerardo Perozziello},
    title = {Microfluidics for Cellular Applications},
    year = {2023},
}

@article{Ookawa2014,
author = {Ookawa, Taiichiro and Inoue, Kazuya and Matsuoka, Makoto and Ebitani, Takeshi and Takarada, Takeshi and Yamamoto, Toshio and Ueda, Tadamasa and Yokoyama, Tadashi and Sugiyama, Chisato and Nakaba, Satoshi and Funada, Ryo and Kato, Hiroshi and Kanekatsu, Motoki and Toyota, Koki and Motobayashi, Takashi and Vazirzanjani, Mehran and Tojo, Seishu and Hirasawa, Tadashi},
doi = {10.1038/srep06567},
issn = {20452322},
journal = {Scientific Reports},
number = {November},
pmid = {25298209},
title = {{Increased lodging resistance in long-culm, low-lignin gh2 rice for improved feed and bioenergy production}},
volume = {4},
year = {2014}
}

@article{Bird2023,
author = {Bird, Rodrigo and Encarna{\c{c}}{\~{a}}o, Lucas},
doi = {10.1016/j.mex.2023.102248},
issn = {2215-0161},
journal = {MethodsX},
number = {102248},
publisher = {Elsevier B.V.},
title = {{Evaluation of the P- $\Delta$ ( P-Delta ) effect in columns and frames using the two-cycle method based on the solution of the beam-column differential equation}},
url = {https://doi.org/10.1016/j.mex.2023.102248},
volume = {11},
year = {2023}
}

@article{Gupta2000,
author = {Gupta, By Akshay and Krawinkler, Helmut},
journal = {Journal of Structural Engineering},
number = {1},
pages = {145--154},
title = {{Dynamic P-delta effects for flexible inelastic steel structures}},
volume = {126},
year = {2000}
}

@article{Gaiotti1989,
author = {Gaiotti, By Regina and Smith, Bryan Stafford},
journal = {Journal of Structural Engineering},
number = {4},
pages = {755--770},
title = {{P-delta analysis of Building Structures}},
volume = {115},
year = {1989}
}

@article{Kawasaki2016,
author = {Kawasaki, Yohei and Tanaka, Yu and Katsura, Keisuke and Purcell, Larry C},
doi = {10.1080/1343943X.2015.1133235},
issn = {1343-943X},
journal = {Plant Production Science},
pages = {1--10},
publisher = {Taylor & Francis},
title = {{Yield and dry matter productivity of Japanese and US soybean cultivars}},
url = {http://dx.doi.org/10.1080/1343943X.2015.1133235},
volume = {1008},
year = {2016}
}

@article{Pinera-Chavez2020,
author = {Pi{\~{n}}era-Chavez, F. J. and Berry, P. M. and Foulkes, M. J. and Molero, G. and Reynolds, M. P.},
doi = {10.1016/j.fcr.2020.107933},
issn = {03784290},
journal = {Field Crops Research},
number = {October 2019},
pages = {107933},
publisher = {Elsevier},
title = {{Optimizing phenotyping methods to evaluate lodging risk for wheat}},
url = {https://doi.org/10.1016/j.fcr.2020.107933},
volume = {258},
year = {2020}
}

@article{Wu2010,
title = {Evaluation of elastic modulus and hardness of crop stalks cell walls by nano-indentation},
journal = {Bioresource Technology},
volume = {101},
number = {8},
pages = {2867-2871},
year = {2010},
issn = {0960-8524},
doi = {https://doi.org/10.1016/j.biortech.2009.10.074},
author = {Yan Wu and Siqun Wang and Dingguo Zhou and Cheng Xing and Yang Zhang and Zhiyong Cai},
}

@article{Xu2020,
author = {Yao Xu and Rui Zhang and Zhaofang Hou and Chao Yan and Xuan Xia and Chunmei Ma and Shoukun Dong and Zhenping Gong},
title = {{Mechanical properties of soybean plants under various plant densities}},
volume = {71},
journal = {Crop and Pasture Science},
number = {3},
publisher = {CSIRO Publishing},
pages = {249 -- 259},
year = {2020},
}

@article{McShane2006,
year = {2006},
month = {aug},
publisher = {},
volume = {16},
number = {10},
pages = {1926},
author = {McShane, G J and Boutchich, M and Srikantha Phani, A and Moore, D F and Lu, T J},
title = {Young's modulus measurement of thin-film materials using micro-cantilevers},
journal = {Journal of Micromechanics and Microengineering},
}

@article{Banks1974, 
title={Anticlastic curvature in anisotropic beams},
volume={78},
number={767}, 
journal={The Aeronautical Journal}, 
author={Banks, P. J.}, 
year={1974}, 
pages={525–528}
}

\end{document}